\documentclass[a4paper]{article}
\usepackage{calc,amsmath,amssymb,amsfonts}
\usepackage[T1]{fontenc}
\usepackage[english]{babel}
\usepackage{xcolor,longfbox,fancyhdr}
\usepackage[top=1in,bottom=0.689in,hmargin=1.25in,nohead,includefoot,foot=0.311in,footskip=0.5827in]{geometry}
\usepackage{enumitem,hyperref}
\hypersetup{colorlinks=true,allcolors=colori,pdfauthor=Li Yang}
\usepackage[pdftex]{graphicx}
\definecolor{colori}{HTML}{0563C1}
\makeatletter\newdimen\@tempdimd\makeatother
\fancypagestyle{Standard}{\fancyhf{}
  \fancyhead[L]{}
  \fancyfoot[C]{\thepage{}}

  \renewcommand\thepage{\arabic{page}}
}
\newcounter{MTSec}

\newcounter{MTChap}

\newcounter{MTEqn}

\providecommand\oiint{\oint}
\author{Li Yang}
\date{2023-08-31}
\begin{document}
\clearpage
\pagestyle{Standard}
{\centering
\textbf{Apply Non-Hermitian Physics to Realize Ultra-High-Quality Factors of Optically Trapped Particles}
\par}

{\centering
Yang Li\textsuperscript{1} and Xiao Li\textsuperscript{1,2}* 
\par}

{\centering
\textsuperscript{1}Department of Physics, Southern University of Science and Technology, \newline
Shenzhen, Guangdong 518055, China
\par}

{\centering
\textsuperscript{2}Department of Physics, The Hong Kong University of Science and Technology, \newline
Hong Kong, China
\par}

{\centering
*E-mail:
\par}

{\centering
Xiao Li: \textcolor[HTML]{0563C1}{lixiao@ust.hk}
\par}

{\centering
Jack Ng: \href{mailto:wuzh3@sustech.edu.cn}{\textcolor[HTML]{0563C1}{wuzh3@sustech.edu.cn}}
\par}

\bigskip

\textbf{Abstract}

Optical trapping and binding systems are non-Hermitian. On one hand, the optical force is non-Hermitian and may pump
energy into the trapped particle when the non-Hermiticity is sufficiently large. On the other hand, the ambient damping
constitutes a loss to the particle. Here, we show that in a low-friction environment\textcolor[HTML]{538135}{,} the
interplay between the energy pumped-in by light and the ambient dissipation can give rise to either instability or a
periodic vibration characterized by a finite quality factor (Q-factor). Through a comprehensive exploration, we analyze
the influence of various parameters on the non-Hermitian force field. Our investigation reveals several strategies for
enhancing the non-Hermitian force field, such as augmenting particle radius and refractive index, utilizing triangular
lattice optical clusters, and reducing lattice constants.

\bigskip

\textbf{Introduction}

Vibrations are almost always accompanied by dissipative damping. In principle, damping can be compensated by gain, but
in most system, it is more easily said than done. We consider optical trapping and binding, where a particle is
confined in vacuum, air, or water. The vibrational modes of such system are subjected to ambient damping, as
consequence, finite Q-factors are expected. Recently, the non-Hermitian characteristics of optical trapping and binding
systems are considered [1]. It is found that by driving the system beyond exceptional point, either by increasing the
nonconservative force or increasing the number of particles, light can pump energy into the system, effectively serving
as a gain. Here, we show that such non-Hermitian gain can indeed compensate the ambient damping in a low-pressure
environment, significantly increases the life-time, or Q-factor, of the modes. 

Optical micromanipulation is a sophisticated technique that utilizes optical forces to trap [2-5], accelerate [6-11],
and pull [12-15] micro- or nano-particles.\textcolor[HTML]{4472C4}{ }The single most widely employed tool for optical
micromanipulation is the optical tweezers [16-18], which employs a tightly focused laser beam to trap a small particle
near the focal point, enabling precise, contact-less, and non-invasive manipulation. Another technique, optical binding
[19], proposed by Burns et al. in 1989, allows for the binding of multiple particles to create a stable lattice
structure known as optical clusters [1, 20, 21] or optical matter [22-24]. Optical binding plays a crucial role in
fields such as colloidal science [21, 25], nanotechnology , [23, 26, 27] and alike [28]. We shall see that the
“non-Hermitian gain” work for both optical trapping and binding in a low-pressure environment.

Aqueous environments have strong dissipation that dominates over many other interactions including the non-Hermitian
gain. In contrast, optical manipulation in vacuum offers some unique advantage due to its low dissipation, for example,
nano-rotors in optical trapping systems can be accelerated to frequencies beyond a GHz [29, 30], making them the
fastest artificial rotors developed to date. Additionally, the unique high quality factor (Q-factor) of the vacuum
environment enables the detection of weak forces and torques [29, 31-33] and facilitates the observation of macroscopic
quantum effects [34-36]. Optical binding can also be used to manipulate multiple particles and create ultrafast
multibody oscillators [37]. These properties make optical micromanipulation in vacuum a promising tool for
ultra-sensitive sensors in high-precision measurement [38, 39]. 

In this study, instead of reducing ambient damping (energy loss), we utilize the non-Hermiticity of optical trapping or
binding to introduce energy gain into particle manipulation and balance the energy loss induced by the unavoidable
ambient damping. Optical micromanipulation is non-Hermitian due to its openness. Despite this non-Hermitian nature,
most of the earlier theoretical studies focused only on conservative forces and ignored nonconservative forces. It is
worth noting that recent studies [40, 41] have shown the importance of non-Hermitian properties in the stability of
trapped or bound particles, with exceptional points (EPs) also being observed [1].\textcolor[HTML]{4472C4}{ }EP emerges
with increasing non-Hermiticity. Before the emergence of EPs (low Hermiticity), the natural vibration frequencies and
the corresponding eigenvectors are numbers. However, after EPs emerge (high non-Hermiticity), a pair of natural
vibration frequencies join togethers to become conjugate pair of complex numbers, simultaneously transforming the
corresponding modes into unstable complex modes. In non-Hermitian optical trapping or binding, the stability of trapped
particles is determined by the competition between ambient damping and the non-Hermitian force field [3, 4, 7, 15]. The
latter pumps energy into the system, allowing the particle to get further away, while the former dissipates the kinetic
energy to keep the particle localize. Here, the particles lose stability when non-Hermiticity is dominant, and vice
versa.

One of the most distinguishing features of the vacuum trapping system is the ultra-high mechanical Q-factor [29, 37, 42]. Here, we propose achieving a very large Q-factor by utilizing non-Hermiticity, which brings energy gain, in optical trapping or binding to balance the energy loss induced by ambient damping. We first examine a single particle trapped by Gaussian beams with different polarizations, and then we study a collection of particles optically bound by several plane waves. We demonstrate that both the optical trapping and binding systems can theoretically achieve an ultra-high Q-factor. Our study exhibits that the non-Hermiticity can be increased in optical binding systems by varying certain parameters, such as assembling triangular lattices and increasing the refractive index and particle radius. The non-Hermiticity of an optical cluster can be characterized by comparing its nonconservative and conservative contributions, which will be presented with respect to different parameters.

\bigskip

\textbf{\textcolor{black}{Results and discussion:}}

\textcolor{black}{The time-averaged optical force is calculated using by}

\begin{equation}
F=\oiint \acute T{\cdot}\mathit{dS},
\label{eq1}
\end{equation}

\textcolor{black}{\ where } $\acute T=\frac 1 2\varepsilon _0EE^{\ast }+\frac 1 2\mu _0HH^{\ast }-\frac 1 4\varepsilon
_0E{\cdot}E^{\ast }\acute I-\frac 1 4\mu _0H{\cdot}H^{\ast }\acute I$ \textcolor{black}{\ is the time-averaged Maxwell
stress tensor, with } $E$ \textcolor{black}{\ and } $H$ \textcolor{black}{\ the total electromagnetic fields. Here, }
$\varepsilon _0$ \textcolor{black}{\ and } $\mu _0$ \textcolor{black}{\ are the permittivity and permeability in free
space, respectively. We consider a collection of spherical particles, whose electromagnetic fields can be obtained
using the multi-particle Mie scattering theory }\textcolor{black}{[20]}\textcolor{black}{.
}\textit{\textcolor{black}{N}}\textcolor{black}{ particles are trapped by optical forces at an equilibrium
configuration, where the force is }\textbf{\textcolor{black}{0}}\textcolor{black}{. For simplicity, we take the
equilibrium configuration to be } $x=0$\textcolor{black}{. The optical force field }
$F\left(x\right)=\left(F_{x,1},F_{y,1},F_{z,1},{\cdots},F_{x,N},F_{y,N},F_{z,N}\right)$ \textcolor{black}{\ can be
linearly approximated as } $F\left(x\right)=\acute K\cdot x$\textcolor{black}{, where }
$x=(x_1,y_1,z_1,{\cdots},x_N,y_N,z_N)$ \textcolor{black}{\ represents the displacement of the particles and } $\acute
K$ \textcolor{black}{\ denotes the force constant matrix, with }
$K_{\mathit{ij}}=\frac{{\partial}F_i}{{\partial}x_j}$\textcolor{black}{\ }\textcolor{black}{[1, 3,
20]}\textcolor{black}{. Here, } $F_i$ \textcolor{black}{\ and } $x_j$ \textcolor{black}{\ denote the
}\textit{\textcolor{black}{i-}}\textcolor{black}{th and }\textit{\textcolor{black}{j-}}\textcolor{black}{th components
of } $F$ \textcolor{black}{\ and } $x$\textcolor{black}{, respectively. } $\acute K$ \textcolor{black}{\ is Hermitian
if } $\acute K^{\dag{}}=\acute K$\textcolor{black}{. However, due to the openness of optical trapping and binding
system, non-Hermitian } $\acute K$\textcolor{black}{, characterized by } $\acute K^{\dag{}}{\neq}\acute
K$\textcolor{black}{, is commonly observed. It is worth noting that, according to Lyapunov's theorem
}\textcolor{black}{[43]}\textcolor{black}{, a nonlinear dynamical system is stable if and only if its linear
approximation (} $F\left(x\right)=\acute K\cdot x$\textcolor{black}{) is stable. Thus the linear system we consider
fully dictates the stability of the particle.}

\textcolor{black}{If one ignores the Brownian fluctuation, which is small at high laser power, the motion of the
particles is governed by}

\begin{equation}
m\frac{d^2x}{dt^2}=\acute K{\bullet}x-\gamma
\frac{dx}{\mathit{dt}},
\label{eq2}
\end{equation}

\textcolor{black}{where } $m$ \textcolor{black}{\ is the mass of a single particle and } $\gamma =m\Gamma _0$
\textcolor{black}{\ is the damping coefficient, with } $\Gamma _0$ \textcolor{black}{\ being inversely proportional to
the environmental press}ure \textit{P} [44-46]. The damping rate  $\Gamma _0$ \ is defined by [47]:

\begin{equation}
\Gamma _0=\frac{6\mathit{\pi \eta r}}
m\frac{0.619}{0.619+\mathit{Kn}}\left(1+c_K\right),
\label{eq3}
\end{equation}
where \textit{$\eta $} is the viscosity of the air,  $\mathit{Kn}=\frac{\lambda _{\mathit{mfp}}} r$ \ is the Knudsen
number,  $\lambda _{\mathit{mfp}}=\frac{68\times 10^{-9}P_{\mathit{atm}}} P$ \ \textcolor{black}{is the mean free path
of air, and } $c_k=\frac{0.31\mathit{Kn}}{\left(0.785+1.152\mathit{Kn}+\mathit{Kn}^2\right)}.$ \textcolor{black}{\ One
can diagonalize } $\acute K$ \textcolor{black}{\ in Eq. }\textcolor{black}{(2)\ to obtain the equation for the
decoupled }\textit{\textcolor{black}{i}}\textcolor{black}{{}-th}\textit{\textcolor{black}{ }}\textcolor{black}{mode:}

\begin{equation}
m\frac{d^2q_i}{dt^2}=K_i{\cdot}q_i-\gamma
\frac{dq_i}{\mathit{dt}},
\label{eq4}
\end{equation}

\textcolor{black}{where } $K_i$ \textcolor{black}{\ is the }\textit{\textcolor{black}{i}}\textcolor{black}{{}-th
eigenvalue of } $\acute K$\textcolor{black}{, and } $q_i$ \textcolor{black}{\ is the
}\textit{\textcolor{black}{i}}\textcolor{black}{{}-th component of the generalized coordinate } $q=\acute
V^{-1}x$\textcolor{black}{. Here, the columns of } $\acute V$ \textcolor{black}{\ are the right-eigenvectors of }
$\acute K$\textcolor{black}{, and one can diagonalize } $\acute K$ \ with  $\acute V^{-1}\acute K\acute
V$\textcolor{black}{. By substituting the particular solution } $q_i=q_{\mathit{i0}}e^{-i\omega _it}$
\textcolor{black}{\ into Eq. }\textcolor{black}{(4), one obtains the vibration frequency for the
}\textit{\textcolor{black}{i}}\textcolor{black}{{}-th mode:}

\begin{equation}
\begin{matrix}\omega _{i,+=\frac{-\mathit{i\gamma }}{2m}+\sqrt{\frac{-K_i} m-\left(\frac{\gamma
}{2m}\right)^2},}\\\omega _{i,-=\frac{-\mathit{i\gamma }}{2m}-\sqrt{\frac{-K_i} m-\left(\frac{\gamma
}{2m}\right)^2}.}\end{matrix}
\label{eq5}
\end{equation}

\textcolor{black}{Here, we adopt a simplified approach by considering vibration frequencies as } $\omega _i=\omega
_{i,+}$ \textcolor{black}{\ within a low-pressure range of 0.1-20 Torr, alongside the corresponding Q factor }
$Q_i=Q_{i,+}$\textcolor{black}{, because the Q factor defined by } $\omega _{i,\pm}$ \textcolor{black}{\ or } $\omega
_{i,-}$ \textcolor{black}{\ is the same for any real } $K_i$ \textcolor{black}{\ when } $\left(\frac{\gamma
}{2m}\right)^2<\frac{-K_i} m$ \textcolor{black}{\ or any complex } $K_i$\textcolor{black}{\ [see Supplementary
Materials]. }

\textcolor{black}{The stability of the particles is essentially determined by } $\Im \left[\omega
_i\right]$\textcolor{black}{, where the time dependence of the mode } $q_i=q_{\mathit{i0}}e^{-i\omega _it}$
\textcolor{black}{\ diverges if } $\Im \left[\omega _i\right]>0$\textcolor{black}{, and vice versa. In between, one
will find a point where } $\Im \left[\omega _i\right]$ \textcolor{black}{\ is exactly 0. In this case, the mode is a
neutral mode. Since it is non-decaying, in this paper, we say that it has a }Q-factor \textcolor{black}{of infinity,
because} \textcolor{black}{[48]}\textcolor{black}{: } $Q_i=\frac{-\left|\Re \left[\omega _i\right]\right|}{2\Im
\left[\omega _i\right]}$\textcolor{black}{. In a perfect vacuum characterized by } $\gamma =0$\textcolor{black}{, }
$Q_i$ \textcolor{black}{\ can also be infinitely large if } $K_i$ \textcolor{black}{\ is real and negative. However, a
perfect vacuum is impossible in an experiment. Here, we utilize the complex } $K_i$ \textcolor{black}{\ induced from
non-Hermiticity (which serve as an effective gain), to compensated the inevitable dissipative losses from the ambient
and realize a large }Q-factor\textcolor{black}{.}

\textcolor{black}{For optical trapping, as depicted in Fig. 1, a single }(\textit{N}=1)\textcolor[HTML]{4472C4}{
}\textcolor{black}{silica particle with a radius of 0.5 $\mu $m and a refractive index of 1.45 is trapped at the
equilibrium position (}\textbf{\textcolor{black}{x=0}}\textcolor{black}{). This trapping is achieved by two
counter-propagating Gaussian beams in the }\textit{\textcolor{black}{z}}\textcolor{black}{{}-direction, each with a
power of 1 mW. Due to the mirror symmetry of the system along the
}\textit{\textcolor{black}{z}}\textcolor{black}{{}-direction, the trapping stiffness in the
}\textit{\textcolor{black}{z}}\textcolor{black}{{}-axis is independent of the transverse
}\textit{\textcolor{black}{xy}}\textcolor{black}{{}-motion }\textcolor{black}{[1]}\textcolor{black}{. Consequently, we
consider a 2-dimensional (2D) force constant matrix, denoted as } $\acute
K_{2D}=\left(\begin{matrix}k_{\mathit{xx}}&k_{\mathit{xy}}\\k_{\mathit{yx}}&k_{\mathit{yy}}\end{matrix}\right)$\textcolor{black}{,
to describe the trapping forces on the transverse plane. All matrix can be split into a symmetric and anti-symmetric
part. By selecting an appropriate coordinate system that diagonalize the symmetric part of } $\acute K_{2D}$
\textcolor{black}{\ one arrives at:}

\begin{equation}
\acute{K'}_{2D}=\left(\begin{matrix}a+b&g\\-g&a-b\end{matrix}\right).
\label{eq6}
\end{equation}

\textcolor{black}{The diagonal elements, } $a+b$\textcolor{black}{, and } $a-b$\textcolor{black}{, represent the
trapping stiffness along the }\textit{\textcolor{black}{x-}}\textcolor{black}{ and
}\textit{\textcolor{black}{y-}}\textcolor{black}{directions, respectively, indicating a two-level distribution in the
frequency spectrum. Here, }\textit{\textcolor{black}{a}}\textcolor{black}{ represents the averaged trapping stiffness,
and }\textit{\textcolor{black}{b}}\textcolor{black}{ represents the half-level spacing between the two vibration
frequency levels. In this setup, the off-diagonal element }\textit{\textcolor{black}{g}}\textcolor{black}{ arises from
the conversion of orbital angular momentum from the spin angular momentum during beam focusing }\textcolor{black}{[1,
49]}\textcolor{black}{. It signifies the non-Hermitian coupling between the trapping stiffness in the
}\textit{\textcolor{black}{x-}}\textcolor{black}{ and }\textit{\textcolor{black}{y-}}\textcolor{black}{directions. The
eigenvalues of } $\acute K_{2D}$ \textcolor{black}{\ can be analytically solved as:}

\begin{equation}
K_{i=1,2}=a\pm \sqrt{b^2-g^2}.
\label{eq7}
\end{equation}

\textcolor{black}{In this expression, an exceptional point (EP) is observed when } $\vee b\vee =\vee g\vee
?$\textcolor{black}{.}

\textcolor{black}{By manipulating the polarization (} $\widehat  p=\widehat  x\cos \left(\xi \right)+i\widehat  y\sin
\left(\xi \right)$\textcolor{black}{) of the input beam, ranging from linear (}\textit{\textcolor{black}{$\xi
$}}\textcolor{black}{ = 0º, Fig. 1}\textbf{\textcolor{black}{a}}\textcolor{black}{) to circular
(}\textit{\textcolor{black}{$\xi $}}\textcolor{black}{ = 45º, Fig. 1}\textbf{\textcolor{black}{b}}\textcolor{black}{),
the corresponding force constant matrix } $\acute K_{2D}$ \textcolor{black}{\ varies with
}\textit{\textcolor{black}{$\xi $}}\textcolor{black}{. Specifically, as }\textit{\textcolor{black}{$\xi
$}}\textcolor{black}{ changes from 0º to 45º }\textcolor{black}{[1]}\textcolor{black}{, } $\vee b\vee ?$
\textcolor{black}{\ decreases to 0 while } $\vee g\vee ?$ \textcolor{black}{\ increases from 0. Therefore, } $\vee
b\vee =\vee g\vee ?$ \textcolor{black}{\ is guaranteed to occur at a certain }\textit{\textcolor{black}{$\xi
$}}\textcolor{black}{, giving rise to the emergence of an exceptional point (EP). In Fig.
1(}\textbf{\textcolor{black}{c}}\textcolor{black}{), the natural vibration frequencies } $\omega _i=\sqrt{\frac{-K_i}
m}$ \textcolor{black}{\ for a particle being trapped in a 2D space are depicted as the polarization varies, where the
ambient damping is neglected. Notably, an EP (green cross) is observed at }\textit{\textcolor{black}{$\xi
$}}\textcolor{black}{ ${\approx}$ 18º. On the left-hand side of the EP, the frequencies } $\omega _i$
\textcolor{black}{\ are purely real, signifying stable particle trapping and an ultra-high }Q-factor\textcolor{black}{
(} $Q_i\rightarrow +{\infty}$\textcolor{black}{) in an ideal vacuum. However, on the right-hand side of the EP, the
frequencies } $\omega _i$ \textcolor{black}{\ become complex conjugates, with one of the } $\Im \left[\omega _i\right]$
\textcolor{black}{\ values being positive. Consequently, the corresponding vibration mode becomes unstable as the time
evolution factor of the vibration amplitude, } $e^{-i\left(\Re \left[\omega _i\right]+i\Im \left[\omega
_i\right]\right)t}=e^{\Im \left[\omega _i\right]t}e^{-i\Re \left[\omega _i\right]t}$\textcolor{black}{, diverges over
time due to the positive } $\Im \left[\omega _i\right]$\textcolor{black}{.}

\bigskip

\lfbox[margin-right=0.0028in,margin-bottom=0.0083in,margin-top=0mm,margin-left=0mm,border-style=none,padding=0mm,vertical-align=top]{\includegraphics[width=5.7681in,height=3.5543in]{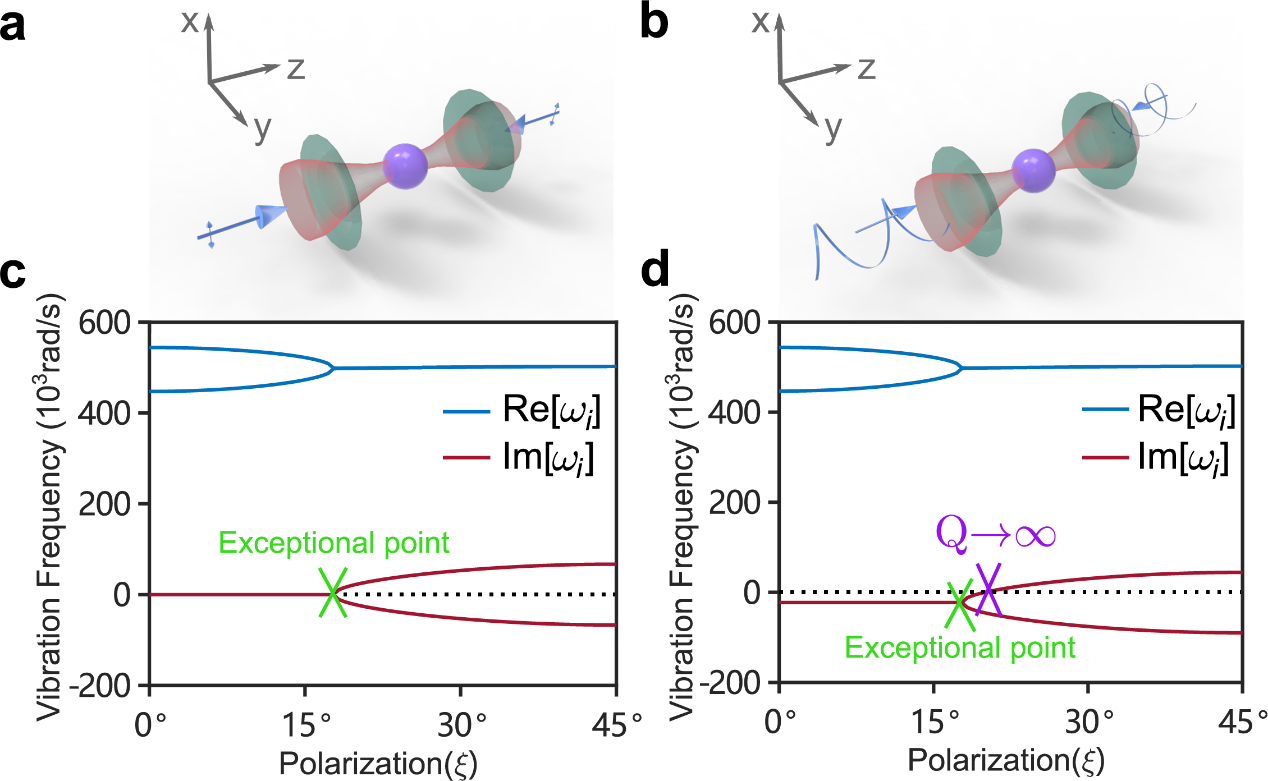}}
\par
\textcolor{black}{Figure 1{\textbar} The vibrational frequencies of an optically trapped silica particle (with a radius
of }\textit{\textcolor{black}{r}}\textcolor{black}{ = 0.5 $\mu $m and refractive index
}\textit{\textcolor{black}{n}}\textcolor{black}{ = 1.45) in a vacuum are investigated as the polarization (} $\widehat 
p=\widehat  x\mathit{cos\xi }+i\widehat  y\mathit{sin\xi }$\textcolor{black}{) of the counter-propagating incident
Gaussian beams varies from linear (}\textit{\textcolor{black}{$\xi $}}\textcolor{black}{ = 0º,
}\textbf{\textcolor{black}{a}}\textcolor{black}{) to circular (}\textit{\textcolor{black}{$\xi $}}\textcolor{black}{ =
45º, }\textbf{\textcolor{black}{b}}\textcolor{black}{) polarization. The incident beams, with a wavelength of
}\textit{\textcolor{black}{$\lambda $}}\textcolor{black}{ = 1.064 $\mu $m, are focused by an objective lens with a
numerical aperture (NA) of 0.9. Each focused beam has a power of 1 mW. The real (blue lines) and imaginary (red lines)
vibrational frequencies in the }\textit{\textcolor{black}{x-y}}\textcolor{black}{ plane are plotted for the particle
trapped in a perfect vacuum at 0 Torr pressure (see (}\textbf{\textcolor{black}{c}}\textcolor{black}{)), as well as in
a low vacuum at 10 Torr pressure (see (}\textbf{\textcolor{black}{d}}\textcolor{black}{)). The exceptional points,
indicated by the green cross in (}\textbf{\textcolor{black}{c}}\textcolor{black}{), mark specific conditions where the
system behavior undergoes a significant change. When the imaginary part of a vibrational frequencies exceeds 0, the
particle loses stability. Additionally, the quality factor (} $Q_i=\frac{-\left|\Re [\omega _i]\right|}{2\Im [\omega
_i]}$\textcolor{black}{) for the vibrational particle approaches infinity } $(+{\infty})$\textcolor{black}{, as denoted
by the purple cross in (}\textbf{\textcolor{black}{d}}\textcolor{black}{), when } $\Im \left[\omega
_i\right]\rightarrow 0$\textcolor{black}{.}

\bigskip

\textcolor{black}{However, ambient damping is inevitable even in a vacuum trapping system and it plays a crucial role in
stabilizing the optically trapped particle, when complex } $K_i$ \textcolor{black}{\ emerges. } $\Re \left[\omega
_i\right]$ \textcolor{black}{\ and } $\Im [\omega _i]$\textcolor{black}{, at a pressure of 10 Torr, are represented by
the blue and red lines in Fig. 1(}\textbf{\textcolor{black}{d}}\textcolor{black}{), respectively. } $\Im [\omega _i]$
\textcolor{black}{\ slightly shifts downwards, resulting in } $\Im \left[\omega _i\right]=0$
\textcolor{black}{\ occurring at $\xi $ ${\approx}$ 20º, indicated by a purple cross, where } $Q\rightarrow
{\infty}$\textcolor{black}{. On the left-hand side of the purple cross, } $\Im \left[\omega _i\right]<0$
\textcolor{black}{\ for all }\textit{\textcolor{black}{i}}\textcolor{black}{, indicating stable particle trapping.
Moreover, it can be noted that ambient damping stabilizes the initially unstable complex } $K_i$
\textcolor{black}{\ modes, which lie between the green (EP) and purple crosses. On the right-hand side of the purple
cross, } $\Im \left[\omega _i\right]$ \textcolor{black}{\ is positive, rendering the particle unstable. However, by
further increasing the ambient damping, we can envision that } $\Im \left[\omega _i\right]$ \textcolor{black}{\ will be
negative for all }\textit{\textcolor{black}{i}}\textcolor{black}{, resulting in particle stability for all
polarizations (}\textit{\textcolor{black}{$\xi $}}\textcolor{black}{).}

\textcolor{black}{In the case of optical binding in a vacuum, as depicted in Figure 2(a), seven dielectric particles are
subjected to binding forces induced by three linearly polarized plane waves propagating at a specific angle in the
}\textit{\textcolor{black}{x-y}}\textcolor{black}{ plane. Each particle has a refractive index of
}\textit{\textcolor{black}{n}}\textcolor{black}{ = 1.1 and a radius of }\textit{\textcolor{black}{r}}\textcolor{black}{
= 0.1 $\mu $m. The intensity (}\textit{I}\textsubscript{0}\textcolor{black}{) of the plane waves is uniformly set at 1
mW/$\mu $m}\textcolor{black}{\textsuperscript{2}}\textcolor{black}{. The particles are bound at the equilibrium
position and form a stable triangular lattice cluster with a lattice constant \~{}} $\sqrt 3\lambda
/2$\textcolor{black}{. The dynamics of this cluster is governed by Eq. }\textcolor{black}{(2). Due to the mirror
symmetry along the }\textit{\textcolor{black}{z}}\textcolor{black}{{}-direction, the motion in the
}\textit{\textcolor{black}{x}}\textcolor{black}{{}-}\textit{\textcolor{black}{y}}\textcolor{black}{ plane is decoupled
from the motion in the }\textit{\textcolor{black}{z}}\textcolor{black}{{}-direction. Therefore, we only focus on the
dynamics within the 2D
}\textit{\textcolor{black}{x}}\textcolor{black}{{}-}\textit{\textcolor{black}{y}}\textcolor{black}{ plane, where the
particle displacement is } $x=\left(x_1,y_1,{\cdots},x_N,y_N\right)$ \textcolor{black}{\ and the force matrix is a
2}\textit{\textcolor{black}{N}}\textcolor{black}{ × 2}\textit{\textcolor{black}{N}}\textcolor{black}{ } $\acute
K$\textcolor{black}{, where }\textit{\textcolor{black}{N}}\textcolor{black}{ = 7 for the case of Fig.
2(}\textbf{\textcolor{black}{a}}\textcolor{black}{). After solving Eq. }\textcolor{black}{(4), one can obtain the
vibrational frequencies } $\omega _i$ \textcolor{black}{\ versus pressure (ambient damping) for each mode, as shown in
Fig. 2(}\textbf{\textcolor{black}{b}}\textcolor{black}{) (} $\Re [\omega _i]$\textcolor{black}{) and Fig.
2(}\textbf{\textcolor{black}{c}}\textcolor{black}{) (} $\Im [\omega _i]$\textcolor{black}{) for the real and imaginary
parts, respectively. }

\lfbox[margin-right=0.0028in,margin-top=0mm,margin-bottom=0mm,margin-left=0mm,border-style=none,padding=0mm,vertical-align=top]{\includegraphics[width=5.7681in,height=1.7736in]{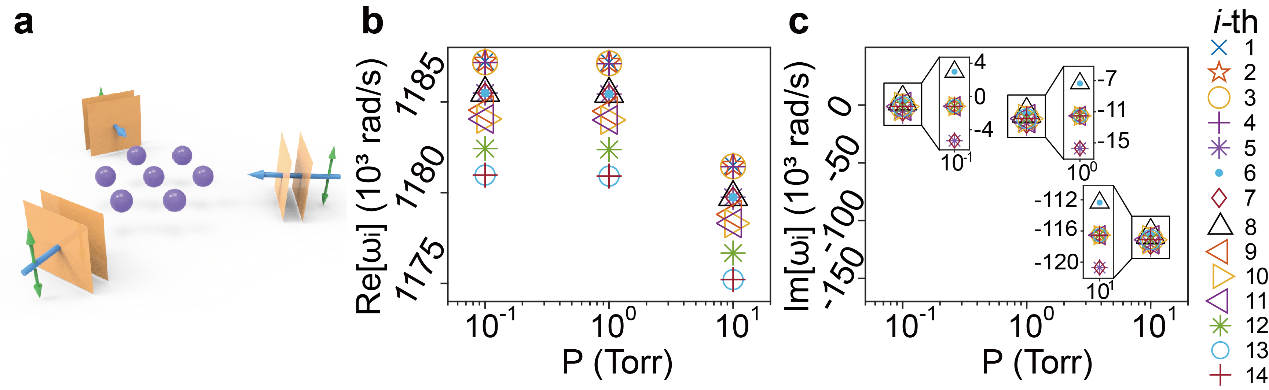}}
\par
Figure 2{\textbar} Vibrational frequencies of an optically bound triangular lattice in a vacuum. (\textbf{a}) The
cluster consists of 7 dielectric particles (\textit{r} = 0.1 $\mu $m, \textit{n} = 1.1) optically bound by 3 linearly
polarized plane waves, each with an intensity (\textit{I}\textsubscript{0}) normalized to 1 mW/$\mu
$m\textsuperscript{2}. As the cluster is bound in a low vacuum with a pressure (\textit{P}) ranging from 0.1 to 10
Torr, the corresponding real (\textbf{b}) and imaginary (\textbf{c}) vibrational frequencies of each vibrational mode
are presented. Here, the particles are considered to move within the \textit{xy}{}-plane.

\bigskip

\textcolor{black}{The symmetry of the lattice structure leads to the degeneracy of certain vibration modes, resulting in
the overlapping vibration frequencies (} $\Re [\omega _i]$\textcolor{black}{) in Fig.
2(}\textbf{\textcolor{black}{b}}\textcolor{black}{). Additionally, the vibration modes associated with complex
conjugate pairs of } $K_i$ \textcolor{black}{\ exhibit two distinct levels of } $\Im \left[\omega
_i\right]$\textcolor{black}{, as depicted in Fig. 2(}\textbf{\textcolor{black}{c}}\textcolor{black}{). According to the
definition of the Q-factor, the mode with the minimum absolute value of {\textbar}} $\Im \left[\omega
_i\right]$\textcolor{black}{{\textbar} corresponds to the maximum Q-factor and is referred to as the maximum Q-factor
(MQF) mode. In Figure 2(}\textbf{\textcolor{black}{c}}\textcolor{black}{), as the pressure
(}\textit{\textcolor{black}{P}}\textcolor{black}{) decreases from 1 Torr to 0.1 Torr, the } $\Im \left[\omega
_{\mathit{MQF}}\right]$ \textcolor{black}{\ becomes positive, indicating the existence of a specific pressure level at
which } $\Im \left[\omega _{\mathit{MQF}}\right]$ \textcolor{black}{\ equals zero, resulting in an ultra-high Q-factor
(} $Q\rightarrow {\infty}$\textcolor{black}{) in optical binding systems.}

\textcolor{black}{According to the definition of the MQF mode, we conducted a comprehensive study on the phase diagram
of MQF in optical binding. In Fig. 3(}\textbf{\textcolor{black}{a}}\textcolor{black}{) and
3(}\textbf{\textcolor{black}{b}}\textcolor{black}{), a triangular lattice (}\textit{\textcolor{black}{N =
}}\textcolor{black}{37, lattice constant \~{}} $\sqrt 3\lambda /2$\textcolor{black}{) and a square lattice
(}\textit{\textcolor{black}{N}}\textcolor{black}{ = 36, lattice constant \~{}} $\lambda $\textcolor{black}{) are
optically bound by three (non-standing wave) and four (standing wave)
}\textit{\textcolor{black}{z}}\textcolor{black}{{}-polarized plane waves, respectively. Different combinations of
particle refractive index (}\textit{\textcolor{black}{n}}\textcolor{black}{) and radius
(}\textit{\textcolor{black}{r}}\textcolor{black}{) were considered: }\textit{\textcolor{black}{n}}\textcolor{black}{ =
1.1, }\textit{\textcolor{black}{r}}\textcolor{black}{ = 0.1 $\mu $m; }\textit{\textcolor{black}{n}}\textcolor{black}{ =
1.2, }\textit{\textcolor{black}{r}}\textcolor{black}{ = 0.1 $\mu $m; }\textit{\textcolor{black}{n}}\textcolor{black}{ =
1.1, }\textit{\textcolor{black}{r}}\textcolor{black}{ = 0.2 $\mu $m; and
}\textit{\textcolor{black}{n}}\textcolor{black}{ = 1.2, }\textit{\textcolor{black}{r}}\textcolor{black}{ = 0.2 $\mu $m.
The phase diagrams of } $1/Q_{\mathit{MQF}}$ \textcolor{black}{\ versus the ambient pressure and plane wave intensity
are shown in Fig. 3(}\textbf{\textcolor{black}{c}}\textcolor{black}{) and
3(}\textbf{\textcolor{black}{d}}\textcolor{black}{) for each combination, where the white regions indicate unstable
optical binding of the particle cluster. In optical binding, non-Hermiticity is influenced not only by the particle
size, refractive index, and illuminating beam intensity but also by non-standing waves, which arise due to its inherent
non-Hermitian nature }\textcolor{black}{[1, 50]}\textcolor{black}{. Consequently, complex modes of } $\acute K$
\textcolor{black}{\ emerge for all combinations of }\textit{\textcolor{black}{n}}\textcolor{black}{ and
}\textit{\textcolor{black}{r}}\textcolor{black}{ in Fig. 3(}\textbf{\textcolor{black}{c}}\textcolor{black}{) when
non-standing waves are present, leading to particle instability (white region) if the ambient damping is insufficient.
On the other hand, in Fig. 3(}\textbf{\textcolor{black}{d}}\textcolor{black}{), complex modes of } $\acute K$
\textcolor{black}{\ only emerge for large }\textit{\textcolor{black}{n}}\textcolor{black}{ and
}\textit{\textcolor{black}{r}}\textcolor{black}{ values (}\textit{\textcolor{black}{n}}\textcolor{black}{ = 1.2,
}\textit{\textcolor{black}{r}}\textcolor{black}{ = 0.2 $\mu $m), where the enhanced non-Hermiticity is attributed to
the high refractive index and particle radius.}

\lfbox[margin-right=0.0028in,margin-bottom=0.002in,margin-top=0mm,margin-left=0mm,border-style=none,padding=0mm,vertical-align=top]{\includegraphics[width=5.7681in,height=4.0398in]{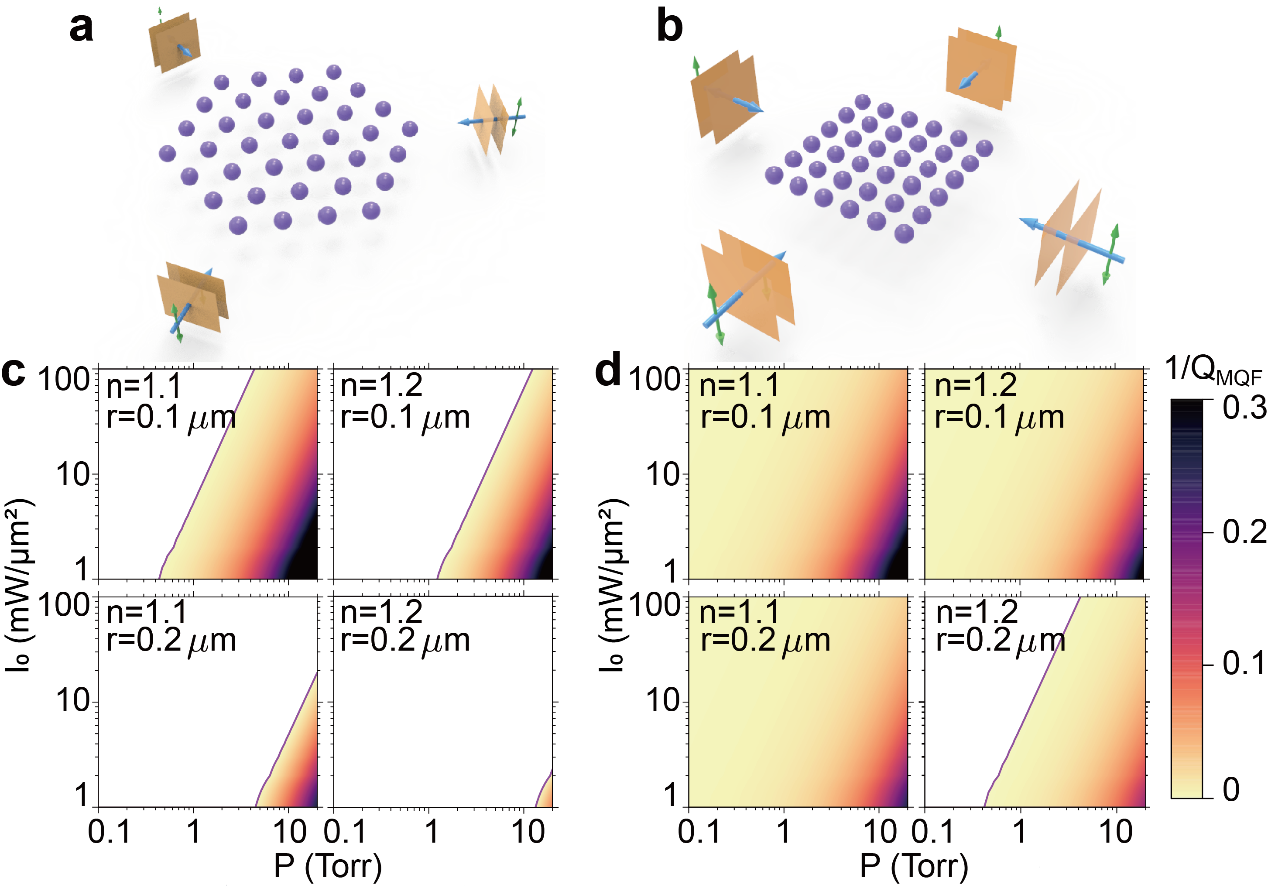}}
\par
\textcolor{black}{Figure 3{\textbar} For optically bound clusters in a vacuum, the phase diagrams of }
$1/Q_{\mathit{MQF}}$ \textcolor{black}{\ versus the pressure (}\textit{\textcolor{black}{P}}\textcolor{black}{) in a
low vacuum range (0.1-20 Torr) and the intensity
(}\textit{\textcolor{black}{I}}\textcolor{black}{\textsubscript{0}}\textcolor{black}{) of each linearly polarized plane
wave (1-100 mW/$\mu $m}\textcolor{black}{\textsuperscript{2}}\textcolor{black}{) are shown for the triangular lattice
(}\textbf{\textcolor{black}{a}}\textcolor{black}{) and square lattice
(}\textbf{\textcolor{black}{b}}\textcolor{black}{). Various combinations of refractive index and particle radius are
considered: }\textit{\textcolor{black}{n }}\textcolor{black}{= 1.1, }\textit{\textcolor{black}{r }}\textcolor{black}{=
0.1 $\mu $m; }\textit{\textcolor{black}{n }}\textcolor{black}{= 1.2, }\textit{\textcolor{black}{r }}\textcolor{black}{=
0.1 $\mu $m; }\textit{\textcolor{black}{n }}\textcolor{black}{= 1.1, }\textit{\textcolor{black}{r }}\textcolor{black}{=
0.2 $\mu $m; and }\textit{\textcolor{black}{n }}\textcolor{black}{= 1.2,
}\textit{\textcolor{black}{r}}\textcolor{black}{ = 0.2 $\mu $m. In panels
(}\textbf{\textcolor{black}{c}}\textcolor{black}{) and (}\textbf{\textcolor{black}{d}}\textcolor{black}{), the pink
lines indicate the ultra-high Q-factors, where } $1/Q_{\mathit{MQF}}\rightarrow 0$\textcolor{black}{.}

\bigskip

\textcolor{black}{In Fig. 3(}\textbf{\textcolor{black}{c-d}}\textcolor{black}{), the pink lines represent the ultra-high
Q-factor (} $1/Q_{\mathit{MQF}}\rightarrow 0$\textcolor{black}{), and they can be accurately fitted with the
logarithmic equation:}

\begin{equation}
\log \left(I_0\right)=a{\cdot}\log\left(P\right)+b.
\label{eq8}
\end{equation}

\textcolor{black}{The slope (} $a$\textcolor{black}{) of Eq. }\textcolor{black}{(8)\ for both the triangular (Fig.
3(}\textbf{\textcolor{black}{c}}\textcolor{black}{)) and square (Fig.
3(}\textbf{\textcolor{black}{d}}\textcolor{black}{)) lattice structures turns out to be a constant value of
}\textit{\textcolor{black}{a}}\textcolor{black}{ = 2, which can be perfectly explained by the definition of the
critical damping coefficient in optical binding. In a vacuum, the non-Hermitian optical binding system exhibits a
critical pressure, characterized by a critical damping coefficient } $\gamma _{\mathit{critical}}=\frac{\sqrt
m\left|\Im \left(K_i\right)\right|}{\sqrt{\Re
\left(K_i\right)}}$\textcolor{black}{\ }\textcolor{black}{[20]}\textcolor{black}{. If the ambient damping (} $\gamma
$\textcolor{black}{) exceeds } $\gamma _{\mathit{critical}}$\textcolor{black}{, the particles remain stable;
conversely, if it is less than } $\gamma _{\mathit{critical}}$\textcolor{black}{, the particles become unstable. This
critical pressure (damping) corresponds to an ultra-high Q-factor, at which } $\Im [\omega _{\mathit{MQF}}]$
\textcolor{black}{\ becomes zero. Since the eigenvalue } $K_i$ \textcolor{black}{\ of the force constant matrix (}
$\acute K$\textcolor{black}{) is proportional to the intensity
(}\textit{\textcolor{black}{I}}\textcolor{black}{\textsubscript{0}}\textcolor{black}{) of the plane wave and the
ambient damping }\textit{\textcolor{black}{$\gamma $}}\textcolor{black}{ is proportional to the ambient pressure
(}\textit{\textcolor{black}{P}}\textcolor{black}{), it is demonstrated that the critical pressure (}
$P_{\mathit{critical}}$\textcolor{black}{) is proportional to the square root of the intensity
(}\textit{\textcolor{black}{I}}\textcolor{black}{\textsubscript{0}}\textcolor{black}{):}

\begin{equation}
P_{\mathit{critical}}{\propto}\sqrt{I_0}.
\label{eq9}
\end{equation}

\textcolor{black}{By taking the logarithm of Eq. }\textcolor{black}{(9)\ and comparing it with Eq.
}\textcolor{black}{(8), we find that the slope }\textit{\textcolor{black}{a}}\textcolor{black}{ = 2. }

\textcolor{black}{Given the significant impact of particle size, refractive index, and standing wave or non-standing
wave on the non-Hermiticity of optically bound particles, we conducted a thorough study of the phase diagrams
illustrating the relationship between the non-Hermiticity and these parameters. By decomposing } $\acute K$
\textcolor{black}{\ into its symmetric (} $\acute S$\textcolor{black}{) and antisymmetric (} $\acute
A$\textcolor{black}{) components, where } $\acute S=\frac{\acute K+\acute K^T} 2$ \textcolor{black}{\ characterizes the
Hermitian (conservative) force field and } $\acute A=\frac{\acute K-\acute K^T} 2$ \textcolor{black}{\ characterizes
the non-Hermitian (nonconservative) force field, we can diagonalize } $\acute S$ \textcolor{black}{\ as }
$\acute{S'}=\acute{\Lambda }\acute S\acute{\Lambda }^{-1}$\textcolor{black}{, with } $\acute{\Lambda }$
\textcolor{black}{\ being the eigenvector matrix of } $\acute S$\textcolor{black}{. Accordingly, } $\acute A$
\textcolor{black}{\ is transformed to } $\acute{A'}=\acute{\Lambda }\acute A\acute{\Lambda }^{-1}$\textcolor{black}{.
We can now define a parameter called } $R$\textcolor{black}{:}

\begin{equation}
R=\frac{{\sum}\left|A_{\mathit{ij}}^{'}\right|}{{\sum}\left|S_{\mathit{ij}}^{'}\right|}.
\label{eq10}
\end{equation}

\textcolor{black}{This parameter quantifies the significance of the non-Hermiticity of } $\acute K$\textcolor{black}{.
Here, } $A_{\mathit{ij}}^{'}$ \textcolor{black}{\ and } $S_{\mathit{ij}}^{'}$ \textcolor{black}{\ are the matrix
elements of } $\acute{A'}$ \textcolor{black}{\ and } $\acute{S'}$\textcolor{black}{, respectively. }

We analyze the phase diagrams of \textit{R} in relation to the particle radius (\textit{r}) and refractive index
(\textit{n}) for various lattice structures. These include triangular lattices consisting of 7 spheres (Fig.
4(\textbf{a1}{}-\textbf{a4})) or 19 spheres (Fig. 4(\textbf{b1}{}-\textbf{b4})), as well as square lattices with 9
spheres (Fig. 4(\textbf{c1}{}-\textbf{c4})) or 16 spheres (Fig. 4(\textbf{d1}{}-\textbf{d4})). We also consider
different lattice constants (\textit{d}) such as  $d=1.1\lambda $\ (Fig. 4(\textbf{b2}{}-\textbf{d2})),  $d=1.5\lambda
$\ (Fig. 4(\textbf{b3}{}-\textbf{d3})), and  $d=2.0\lambda $\ (Fig. 4(\textbf{b4}{}-\textbf{d4})). In Figure 4, it is
observed that optically bound clusters induced by non-standing waves (triangular lattice) exhibit larger values of
\textit{R} compared to those induced by standing waves (square lattice). The phase diagrams clearly demonstrate the
significant influence of non-standing waves, which is non-Hermitian (non-conservative) by nature, on \textit{R}.
Increasing the values of \textit{r}, \textit{n}, or the number of particles (\textit{N}) enhances multiple scattering,
resulting in larger values of \textit{R} in the phase diagrams. Conversely, increasing the lattice constant
(\textit{d}) weakens multiple scattering, leading to a decrease in \textit{R} across the phase diagrams. The highest
impact of non-Hermiticity is observed in Fig. 4(\textbf{b2}), while the lowest impact is observed in Fig.
4(\textbf{c4}). Furthermore, it is important to note that \textit{R} is also influenced \textcolor{black}{by the
optical path difference, characterized by } $2r\left(n-1\right)=\mathit{const}$\textcolor{black}{, between the light
passing through a single particle and a vacuum, as observed in the hyperbolic patterns within} the phase diagrams. For
instance, the hyperbolic pattern in Fig. 4(\textbf{b3}) can be accurately described by the equation 
$2r\left(n-1\right)=0.7\lambda $, as indicated by the pink line in Fig. 4(\textbf{b3}).

\lfbox[margin-right=0.0028in,margin-bottom=0in,margin-top=0in,margin-left=0mm,border-style=none,padding=5mm,vertical-align=top]{\includegraphics[width=5.7681in,height=5.7681in]{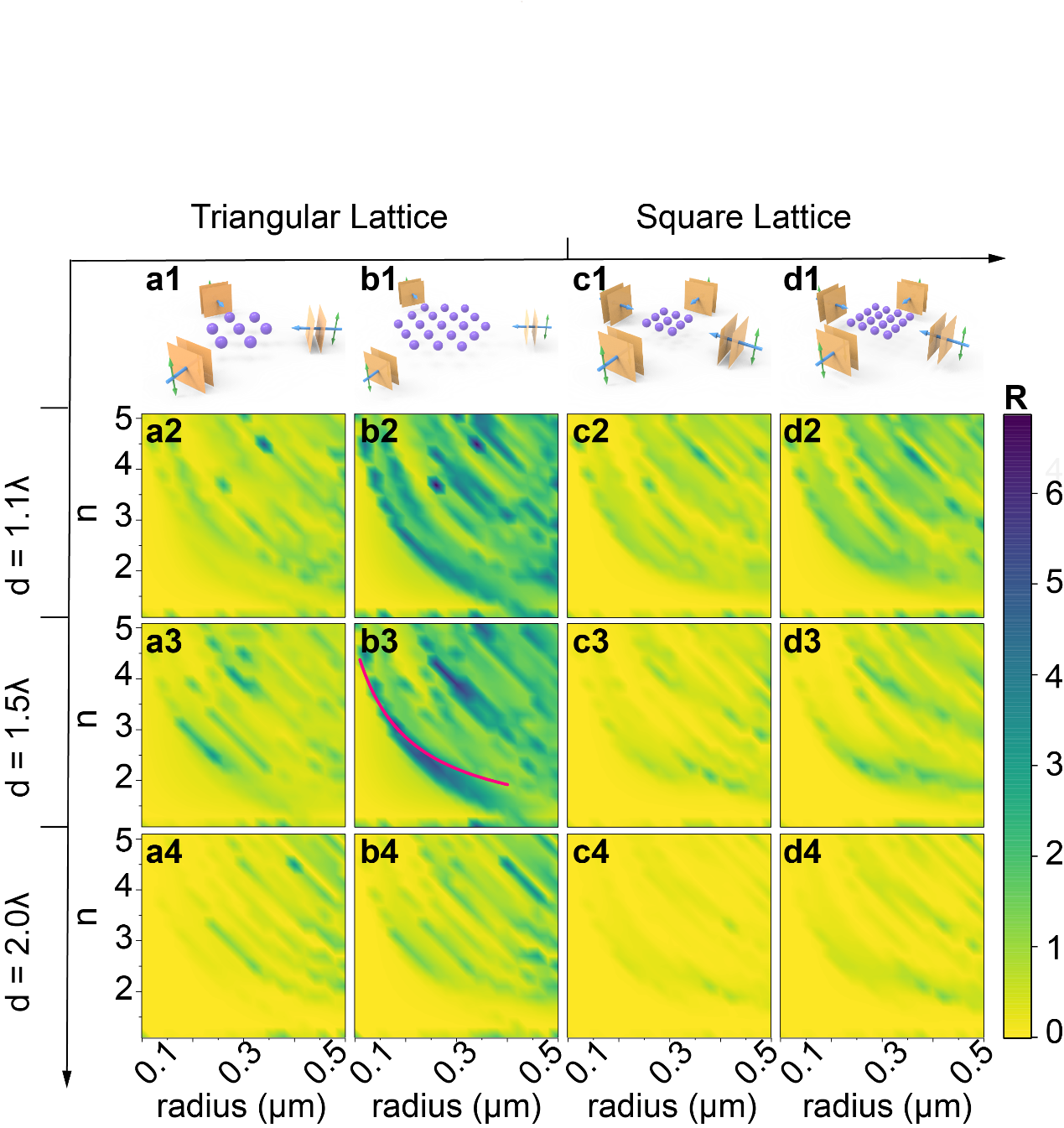}}
\par
\textcolor{black}{Figure 4{\textbar} The phase diagrams depict the non-Hermiticity of optically bound clusters in a
vacuum for triangular lattice (}\textbf{\textcolor{black}{a1}}\textcolor{black}{ and
}\textbf{\textcolor{black}{b1}}\textcolor{black}{) and square lattice
(}\textbf{\textcolor{black}{c1}}\textcolor{black}{ and}\textbf{\textcolor{black}{ d1}}\textcolor{black}{), versus the
radius }\textit{\textcolor{black}{r}}\textcolor{black}{ (from 0.1 $\mu $m to 0.5 $\mu $m) and refractive index
}\textit{\textcolor{black}{n}}\textcolor{black}{ (from 0.1 $\mu $m to 0.5 $\mu $m) of the particles. The
non-Hermiticity is characterized by the}\textit{\textcolor{black}{ }}
$R=\frac{{\sum}\left|A_{\mathit{ij}}^{'}\right|}{{\sum}\left|S_{\mathit{ij}}^{'}\right|}$\textcolor{black}{. The phase
diagrams cover various lattice structures, including a triangular lattice with 7 particles
(}\textbf{\textcolor{black}{a2}}\textcolor{black}{{}-}\textbf{\textcolor{black}{a4}}\textcolor{black}{), a triangular
lattice with 19 particles
(}\textbf{\textcolor{black}{b2}}\textcolor{black}{{}-}\textbf{\textcolor{black}{b4}}\textcolor{black}{), a square
lattice with 9 particles
(}\textbf{\textcolor{black}{c2}}\textcolor{black}{{}-}\textbf{\textcolor{black}{c4}}\textcolor{black}{), and a square
lattice with 16 particles
(}\textbf{\textcolor{black}{d2}}\textcolor{black}{{}-}\textbf{\textcolor{black}{d4}}\textcolor{black}{). Additionally,
different lattice constants are considered, namely } $d=1.1\lambda
$\textcolor{black}{\ (}\textbf{\textcolor{black}{a2-d2}}\textcolor{black}{); } $d=1.5\lambda
$\textcolor{black}{\ (}\textbf{\textcolor{black}{a3-d3}}\textcolor{black}{); } $d=2.0\lambda
$\textcolor{black}{\ (}\textbf{\textcolor{black}{a4-d4}}\textcolor{black}{). The diagrams reveal distinctive hyperbolic
patterns resulting from the interference structure characterized by }
$2r\left(n-1\right)=\mathit{const}$\textcolor{black}{. For instance, one of the hyperbolic curves in
(}\textbf{\textcolor{black}{b3}}\textcolor{black}{) can be accurately fitted with a pink line representing 2r(n-1) =
0.7$\lambda $.}

\bigskip

\textbf{\textcolor{black}{Conclusions }}

\textcolor{black}{In conclusion, non-Hermitian physics is applied to achieve an ultra-high Q-factor for an optically
trapped particle or an optically bound multi-particle lattice in a vacuum, accounting for the presence of ambient
damping. By adjusting the non-Hermiticity or ambient damping, it becomes possible to achieve a very high Q-factor. The
parameters influencing non-Hermiticity in optical binding have been comprehensively discussed and presented. These
ultra-high Q-factor oscillators have significant potential in the measurement of weak forces and torques
}\textcolor{black}{[31, 38, 39, 51]}\textcolor{black}{, given that measurement sensitivity is inversely proportional to
the Q-factor }\textcolor{black}{[28, 29]}\textcolor{black}{.}

\bigskip

\bigskip

\clearpage
\textbf{Reference:}

1.\ \ Li, X., et al., \textit{Non-Hermitian physics for optical manipulation uncovers inherent instability of large
clusters.} Nature Communications, 2021. \textbf{12}(1): p. 6597.

2.\ \ Ashkin, A., \textit{History of optical trapping and manipulation of small-neutral particle, atoms, and molecules.}
IEEE Journal of Selected Topics in Quantum Electronics, 2000. \textbf{6}(6): p. 841-856.

3.\ \ Ng, J., Z. Lin, and C.T. Chan, \textit{Theory of Optical Trapping by an Optical Vortex Beam.} Physical Review
Letters, 2010. \textbf{104}(10): p. 103601.

4.\ \ Li, Y., L.-M. Zhou, and N. Zhao, \textit{Anomalous motion of a particle levitated by Laguerre–Gaussian beams.}
Optics Letters, 2021. \textbf{46}(1): p. 106-109.

5.\ \ Yuanjie, Y., et al., \textit{Optical trapping with structured light: a review.} Advanced Photonics, 2021.
\textbf{3}(3): p. 034001.

6.\ \ Arita, Y., et al., \textit{Coherent oscillations of a levitated birefringent microsphere in vacuum driven by
nonconservative rotation-translation coupling.} Science Advances. \textbf{6}(23): p. eaaz9858.

7.\ \ Shao, L., et al., \textit{Gold Nanorod Rotary Motors Driven by Resonant Light Scattering.} ACS Nano, 2015.
\textbf{9}(12): p. 12542-12551.

8.\ \ Shao, L. and M. Käll, \textit{Light-Driven Rotation of Plasmonic Nanomotors.} Advanced Functional Materials, 2018.
\textbf{28}(25): p. 1706272.

9.\ \ Tang, S., et al., \textit{Structure-Dependent Optical Modulation of Propulsion and Collective Behavior of
Acoustic/Light-Driven Hybrid Microbowls.} Advanced Functional Materials, 2019. \textbf{29}(23): p. 1809003.

10.\ \ Li, Y., et al., \textit{Plasmon-coupling-induced photon scattering torque.} Journal of the Optical Society of
America B, 2022. \textbf{39}(3): p. 671-676.

11.\ \ Wu, X., et al., \textit{Light-driven microdrones.} Nature Nanotechnology, 2022. \textbf{17}(5): p. 477-484.

12.\ \ Chen, J., et al., \textit{Optical pulling force.} Nature Photonics, 2011. \textbf{5}(9): p. 531-534.

13.\ \ Ding, K., et al., \textit{Realization of optical pulling forces using chirality.} Physical Review A, 2014.
\textbf{89}(6): p. 063825.

14.\ \ Wang, H., et al., \textit{Janus Magneto–Electric Nanosphere Dimers Exhibiting Unidirectional Visible Light
Scattering and Strong Electromagnetic Field Enhancement.} ACS Nano, 2015. \textbf{9}(1): p. 436-448.

15.\ \ Li, X., et al., \textit{Optical pulling at macroscopic distances.} Science Advances, 2019. \textbf{5}(3): p.
eaau7814.

16.\ \ Ashkin, A., \textit{Acceleration and Trapping of Particles by Radiation Pressure.} Physical Review Letters, 1970.
\textbf{24}(4): p. 156-159.

17.\ \ Ashkin, A. and J.M. Dziedzic, \textit{Optical levitation in high vacuum.} Applied Physics Letters, 1976.
\textbf{28}(6): p. 333-335.

18.\ \ Ashkin, A., et al., \textit{Observation of a single-beam gradient force optical trap for dielectric particles.}
Optics Letters, 1986. \textbf{11}(5): p. 288-290.

19.\ \ Burns, M.M., J.-M. Fournier, and J.A. Golovchenko, \textit{Optical binding.} Physical Review Letters, 1989.
\textbf{63}(12): p. 1233-1236.

20.\ \ Ng, J., et al., \textit{Photonic clusters formed by dielectric microspheres: Numerical simulations.} Physical
Review B, 2005. \textbf{72}(8): p. 085130.

21.\ \ Nan, F., et al., \textit{Dissipative Self-Assembly of Anisotropic Nanoparticle Chains with Combined
Electrodynamic and Electrostatic Interactions.} Advanced Materials, 2018. \textbf{30}(45): p. 1803238.

22.\ \ Burns, M.M., J.-M. Fournier, and J.A. Golovchenko, \textit{Optical Matter: Crystallization and Binding in Intense
Optical Fields.} Science, 1990. \textbf{249}(4970): p. 749-754.

23.\ \ Parker, J., et al., \textit{Optical matter machines: angular momentum conversion by collective modes in optically
bound nanoparticle arrays.} Optica, 2020. \textbf{7}(10): p. 1341-1348.

24.\ \ Huang, C.-H., et al., \textit{The primeval optical evolving matter by optical binding inside and outside the
photon beam.} Nature Communications, 2022. \textbf{13}(1): p. 5325.

25.\ \ Wei, M.-T., et al., \textit{Lateral optical binding between two colloidal particles.} Scientific Reports, 2016.
\textbf{6}(1): p. 38883.

26.\ \ Tanaka, Y.Y., et al., \textit{Plasmonic linear nanomotor using lateral optical forces.} Science Advances, 2020.
\textbf{6}(45): p. eabc3726.

27.\ \ Duan, X.-Y., et al., \textit{Transverse optical binding for a dual dipolar dielectric nanoparticle dimer.}
Physical Review A, 2021. \textbf{103}(1): p. 013721.

28.\ \ Svak, V., et al., \textit{Stochastic dynamics of optically bound matter levitated in vacuum.} Optica, 2021.
\textbf{8}(2): p. 220-229.

29.\ \ Ahn, J., et al., \textit{Ultrasensitive torque detection with an optically levitated nanorotor.} Nature
Nanotechnology, 2020. \textbf{15}(2): p. 89-93.

30.\ \ Jin, Y., et al., \textit{6 GHz hyperfast rotation of an optically levitated nanoparticle in vacuum.} Photonics
Research, 2021. \textbf{9}(7): p. 1344-1350.

31.\ \ Jain, V., et al., \textit{Direct Measurement of Photon Recoil from a Levitated Nanoparticle.} Physical Review
Letters, 2016. \textbf{116}(24): p. 243601.

32.\ \ Ahn, J., et al., \textit{Optically Levitated Nanodumbbell Torsion Balance and GHz Nanomechanical Rotor.} Physical
Review Letters, 2018. \textbf{121}(3): p. 033603.

33.\ \ Nan, F. and Z. Yan, \textit{Optical Sorting at the Single-Particle Level with Single-Nanometer Precision Using
Coordinated Intensity and Phase Gradient Forces.} ACS Nano, 2020. \textbf{14}(6): p. 7602-7609.

34.\ \ Tebbenjohanns, F., et al., \textit{Quantum control of a nanoparticle optically levitated in cryogenic free
space.} Nature, 2021. \textbf{595}(7867): p. 378-382.

35.\ \ Zhang, H., X. Chen, and Z.-q. Yin, \textit{Quantum Information Processing and Precision Measurement Using a
Levitated Nanodiamond.} Advanced Quantum Technologies, 2021. \textbf{4}(8): p. 2000154.

36.\ \ Shen, K., et al., \textit{On-chip optical levitation with a metalens in vacuum.} Optica, 2021. \textbf{8}(11): p.
1359-1362.

37.\ \ Arita, Y., E.M. Wright, and K. Dholakia, \textit{Optical binding of two cooled micro-gyroscopes levitated in
vacuum.} Optica, 2018. \textbf{5}(8): p. 910-917.

38.\ \ Ranjit, G., et al., \textit{Zeptonewton force sensing with nanospheres in an optical lattice.} Physical Review A,
2016. \textbf{93}(5): p. 053801.

39.\ \ Shi, H., et al., \textit{Optical binding and lateral forces on chiral particles in linearly polarized plane
waves.} Physical Review A, 2020. \textbf{101}(4): p. 043808.

40.\ \ Feng, L., R. El-Ganainy, and L. Ge, \textit{Non-Hermitian photonics based on parity–time symmetry.} Nature
Photonics, 2017. \textbf{11}(12): p. 752-762.

41.\ \ Wang, H., et al., \textit{Topological physics of non-Hermitian optics and photonics: a review.} Journal of
Optics, 2021. \textbf{23}(12): p. 123001.

42.\ \ Zhou, L.-M., et al., \textit{Optical levitation of nanodiamonds by doughnut beams in vacuum.} Laser \& Photonics
Reviews, 2017. \textbf{11}(2): p. 1600284.

43.\ \ Seyranian, A.P. and A.A. Mailybaev, \textit{Multiparameter stability theory with mechanical applications}. Vol.
13. 2003: World Scientific.

44.\ \ Beresnev, S.A., V.G. Chernyak, and G.A. Fomyagin, \textit{Motion of a spherical particle in a rarefied gas. Part
2. Drag and thermal polarization.} Journal of Fluid Mechanics, 1990. \textbf{219}: p. 405-421.

45.\ \ Li, T., S. Kheifets, and M.G. Raizen, \textit{Millikelvin cooling of an optically trapped microsphere in vacuum.}
Nature Physics, 2011. \textbf{7}(7): p. 527-530.

46.\ \ Hempston, D., et al., \textit{Force sensing with an optically levitated charged nanoparticle.} Applied Physics
Letters, 2017. \textbf{111}(13): p. 133111.

47.\ \ Gieseler, J., et al., \textit{Subkelvin Parametric Feedback Cooling of a Laser-Trapped Nanoparticle.} Physical
Review Letters, 2012. \textbf{109}(10): p. 103603.

48.\ \ Jackson, J.D., \textit{Classical Electrodynamics}. 3rd ed. 1999: Wiley. 371-372.

49.\ \ Zhao, Y., et al., \textit{Spin-to-Orbital Angular Momentum Conversion in a Strongly Focused Optical Beam.}
Physical Review Letters, 2007. \textbf{99}(7): p. 073901.

50.\ \ Du, J., et al., \textit{Tailoring Optical Gradient Force and Optical Scattering and Absorption Force.} Scientific
Reports, 2017. \textbf{7}(1): p. 18042.

51.\ \ Metzger, N.K., et al., \textit{Measurement of the Restoring Forces Acting on Two Optically Bound Particles from
Normal Mode Correlations.} Physical Review Letters, 2007. \textbf{98}(6): p. 068102.

\bigskip
\end{document}